\documentclass[10pt, conference]{IEEEtran}
 
\usepackage{cite}
\usepackage{amsmath,amssymb,amsfonts}
\usepackage{graphicx}
\usepackage{textcomp}
\usepackage{xcolor}
\def\BibTeX{{\rm B\kern-.05em{\sc i\kern-.025em b}\kern-.08em
    T\kern-.1667em\lower.7ex\hbox{E}\kern-.125emX}}

\usepackage{url}
\usepackage{csquotes}
\usepackage{underscore}
\usepackage{algpseudocode}
\newtheorem{problem}{Problem}

\begin{document}

\title{Mining Program Properties From Neural Networks Trained on Source Code Embeddings}

\author{\IEEEauthorblockN{Martina Saletta, Claudio Ferretti}
\IEEEauthorblockA{\textit{Department of Informatics, Systems and Communication} \\
\textit{University of Milano-Bicocca}\\
Milan, Italy \\
martina.saletta@unimib.it, claudio.ferretti@unimib.it}
}


\maketitle

\begin{abstract}
In this paper, we propose a novel approach for mining different program features by analysing the internal behaviour of a deep neural network trained on source code. 

Using an unlabelled dataset of Java programs and three different embedding strategies for the methods in the dataset, we train an autoencoder for each program embedding and then we test the emerging ability of the internal neurons in autonomously building internal representations for different program features. We defined three binary classification labelling policies inspired by real programming issues, so to test the performance of each neuron in classifying programs accordingly to these classification rules, showing that some neurons can actually detect different program properties.

We also analyse how the program representation chosen as input affects the performance on the aforementioned tasks. On the other hand, we are interested in finding the overall most informative neurons in the network regardless of a given task. To this aim, we propose and evaluate two methods for ranking neurons independently of any property. 

Finally, we discuss how these ideas can be applied in different settings for simplifying the programmers' work, for instance if included in environments such as software repositories or code editors.  

\end{abstract}


\section{Introduction}

Many research results on the use of deep learning systems show how the internal representation developed by a system during its training is of value, even for tasks different than that it was trained for. Consider, for example, the methods for pretraining \cite{erhan2010does} or semisupervised learning, transfer learning \cite{zhuang2020comprehensive}, internal interpretability \cite{gilpin2018explaining}.
The study of these approaches is active for the domains of image processing (e.g.\cite{le:gatti}) and of natural language processing (NLP) \cite{dalvi:sabbia}, while fewer results are, however, available for the domain of source code processing.

The promise of this research is that we can develop systems based on neural learning by assembling modules, for instance an embedding module as first input stage, a supervised system for an ancillary task, but stripped of its output layer, and finally a different output layer, trained on the main task.

Going further with this approach, namely the decomposition and reassembling of parts from neural systems, we can consider to check whether internal neurons, during the training of the whole network, have developed some capability of performing useful classifications of input instances, as defined by their activation level for each input.

In the field of program comprehension, many neural systems have been presented (we refer the reader to \cite{triet:sourcecodemodelling} and \cite{allamanis:mlbigcode} for surveys), for tasks relevant to the field. Furthermore, some results are known about ways for embedding source code in vector spaces. We consider interesting to actively pursue, in this specific context of source code embeddings, the goal of being able to examine internal neurons of some given learning system, in order to search for those which exhibit interesting behaviours in terms of classification performance or activation patterns.

Moreover, with neural systems for program comprehension becoming bigger and more sophisticated in their architecture, more and more useful neurons could be mined out from their internal structure. Specific to this context, is the need for embedding systems of being effective for source code-related tasks, but also the definition of what is ``useful'' deserves attention. For instance, the design of models that can be used for easing the effort of developers working on source code, e.g. tagging code fragments or indexing code lines for semantic searches, is of interest in the program comprehension field. And, eventually, the outcome would be to have tools which embed in themselves neural models for exploring source code that are trained for a main task on it, but which also allow to exploit, for other needs of the developer, further parts of the internal representation developed while being trained.
We could conceive this perspective, in the long term, as a way to develop a versatile assistant for people working on source code.

In this work, we present the following contributions:
\begin{itemize}
    \item we evaluate how three different embedding methods for source code perform when used to feed a simple unsupervised system by looking for internal neurons which are good in chosen code classification tasks, aimed at simple properties, both structural and semantic;
    \item we define a scoring function to identify internal neurons which are able to discriminate methods, from an information theoretic point of view, which is not linked to specific classification tasks;
    \item we test our scoring function, together with one from literature, in a simple system, succeeding in finding neurons which exhibit some capabilities to cluster specific groups of methods from the thousands seen as input;
    \item finally, we discuss how this approach could be useful, with the growth of neural-based tools for program comprehension, for future application.
\end{itemize}

\section{Background and Related Work}

The overall goal is to enable a system to continuously help the developer by examining the code he is working on by highlighting to him code fragments enjoining a useful set of properties. Our approach is to first map the source code to feature vectors, for instance via a neural embedding, then to analyse the internal behaviour of an unsupervised neural network trained on such vectors, and to finally look for neurons which are good for code classification tasks.

For the first step, that is mapping source code blocks to vectors, the methods known in literature range from building the vectors from identifiers and data types found in code \cite{yamaguchi:vulnDiscovery}, to neural embeddings trying to keep information about syntax trees and call/control graphs \cite{mou:tbcnn,allamanis:graphs}.

Usually, these methods are defined while keeping into account some specific downstream task, for instance to help program comprehension with code search \cite{gu:search,ling:search}.
Instead, we are interested in defining, or in evaluating, embeddings used to feed an unsupervised learning system, and jointly to study how the system trained in this way develops useful internal representations.

In the area of image analysis, a similar approach was devised in \cite{le:gatti}, where each neuron of an unsupervised trained network was evaluated with respect to a given image classification task, with insightful results.
More recently, results have been obtained for evaluating single neurons for sentiment analysis tasks \cite{radford:reviews}, or in networks trained to model natural languages \cite{dalvi:sabbia}. In the following, we also will evaluate this latter method.

Our aim is to help bridging the gap between the availability of neural systems which develop useful internal representation of source code, for instance when trained for searching code with specific characteristics in large codebases \cite{shuai:search}, and the development of other tools helping programmers (for instance for tasks like those in \cite{li:tagging,zhang:bugs}). We bring to the domain of source code data sets the aforementioned approach, that of evaluating single neurons inside the trained systems, by defining a new scoring function, and then we test this on a simple unsupervised neural system.

The perspective is to fruitfully apply it to bigger trained systems, in order of being able to discover and to use all the internally learned knowledge, loosely inspired by what already happens in human brain: it is known that experts in specific fields develops further unconscious perceptual capabilities, driven by their exercise on the main task of their expertise \cite{kiesel2009playing}.

\section{Experimental Settings}

This section describes the experimental setting we adopted in order to study the ability of the hidden neurons of a generic neural network in building an high-level representation for some specific source code-related features starting from different program vector representations, similarly to what the authors of~\cite{le:gatti} and~\cite{radford:reviews} did for images and natural language, respectively.

We first trained three simple autoencoders (i.e. artificial neural networks trained in the reconstruction of the input, see~\cite{goodfellow:deepLearning} for a complete reference) on three different source code embeddings, and then we performed two categories of experiments:

\begin{enumerate}
    \item Binary classification experiments for ranking neurons considering their ability in solving specific tasks.
    \item Analysis of the relevance of the neurons for the network itself, regardless of a given task.
\end{enumerate}

\subsection{Network Architecture}

\begin{figure}
    \centering
    \includegraphics[width=\linewidth]{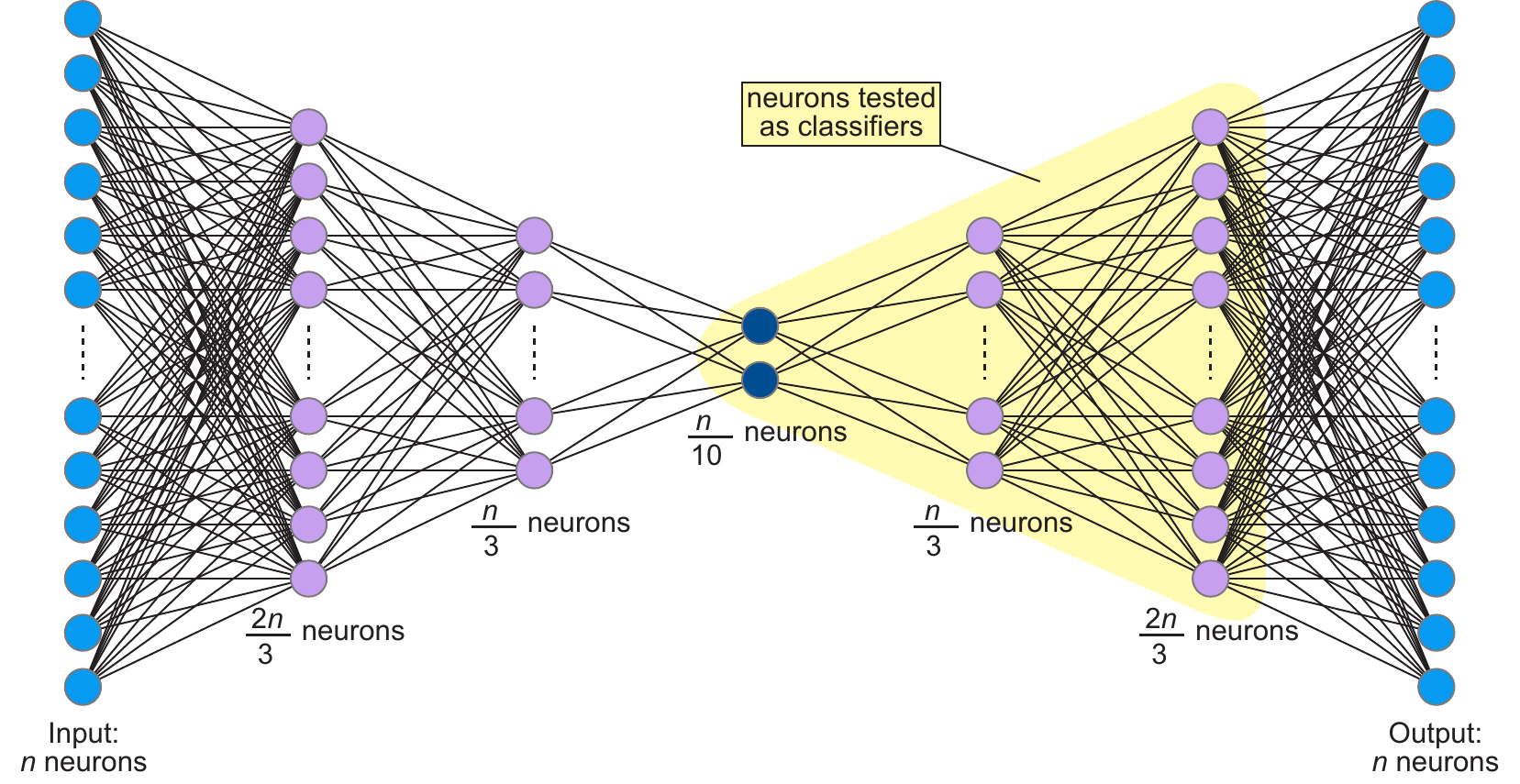}
    \caption{Network architecture. The sizes of the layers differs in the models due to different embedding dimensions: we considered $n=384$ for code2vec vectors and $n=300$ for doc2vec and ast2vec vectors. Layers highlighted in yellow are those tested in the experiments.}
    \label{fig:autoencoder}
\end{figure}

Since the aim of this work is to investigate the potential of a new methodological approach for mining varied properties from a program, the study of the best network architecture is not in the scope of this paper. Therefore, without any hyper-parameters optimization nor any in-depth study on the network design, we implemented three simple dense autoencoders having two hidden layers in the encoder, two symmetrical hidden layers in the decoder and one code layer in the middle, as shown in Figure~\ref{fig:autoencoder}. We used the \emph{ADADELTA} optimizer~\cite{zeiler:adadelta} and the Mean Squared Error as a loss function. As the activation function for the hidden layers we applied the Rectified Linear Unit (ReLU) defined, for all $x \in \mathbb{R}$ as $\max(0, x)$. All the models are implemented using the APIs provided by the Keras library~\cite{chollet:keras}.

\subsection{Training}
We trained our models on the dataset used in~\cite{allamanis:summarization}, which is a collection of popular GitHub\footnote{\url{https://github.com}} Java projects that contains over 400000 methods. The dataset is already split into training, test and validation sets. We used a random $75$:$25$ split of the methods in the training set for the training phase of the autoencoders and samples of the methods in the union of the test and validation sets for the experiments.

For each method in the dataset, we computed a corresponding vector representation using three different source code embedding algorithms:

\begin{enumerate}
    \item The source code embedding obtained with the code2vec model\cite{alon:code2vec}, which produces $384$-dimensional vectors by combining the paths on the abstract syntax tree (AST) through the attention mechanism~\cite{Dbahdanau:attention}.
    
    \item A $300$-dimensional embedding obtained by simply applying the doc2vec model~\cite{le:doc2vec} to a pretty printed version of the methods in the dataset using the gensim framework~\cite{gensim}.
    
    \item The source code embedding proposed in \cite{saletta:ast2vec}, which we will refer to as ast2vec, consisting in the application of the word2vec model~\cite{mikolov:word2vec} to words and sentences derived from the abstract syntax tree. 
\end{enumerate}

The reason of the choice of such embeddings lies in the main purpose of this work of investigating the potential of our approach for mining diverse program properties. Therefore, we tested it on program vectors having different underlying construction ideas: assuming that the name a programmer gives to a method is related to its functionality, and since code2vec is originally designed for predicting that name, code2vec vectors are expected to hold some semantic information related to the method they represent; doc2vec embedding is built by applying a classical NLP technique to the pure source code, so it is not supposed to be particularly viable in this context, while ast2vec is developed by considering both structural (derived from the AST) and lexical features, and thus it is assumed to be more flexible for general program comprehension applications than code2vec and doc2vec.

We trained each autoencoder for $50$ epochs using the same hyper-parameters except for the layer dimensions, defined as a function of the input dimension $n$ (see Figure~\ref{fig:autoencoder}), which is set to $n=384$ for code2vec and $n=300$ for ast2vec and doc2vec.

\section{Classification Experiments}\label{sec:classification}

For assessing the ability of the internal neurons in our networks to build internal representations for different program properties, we first tested each neuron in being used as a classifier for different binary classification problems, following the same approach of previous works (see~\cite{le:gatti} and~\cite{radford:reviews}). 

\subsection{Problems definition}

Dealing with images and product reviews as in the referred papers, the properties according to which to classify the input objects can be easily defined: could be, for instance, the presence of particular patterns (e.g. cats or faces \cite{le:gatti}) or positive and negative review sentiments~\cite{radford:reviews}, as in classical image recognition and sentiment analysis tasks. In the program comprehension context, however, such kind of properties do not directly arise from the source code, or at least are not immediately evident for a human being reading it. Therefore, we first defined three labelling policies for classification, so to capture properties having different natures. The first one, designed using the Control Flow Graph (CFG)~\cite{allen:cfg}, addresses the syntactical structure of a method in terms of its structural complexity, the second one relies on the method's identifiers chosen by the programmers in order to target a semantic task, while the last one is a random labelling strategy used as a baseline.    

\paragraph{Structural labelling policy} 
We consider the \emph{cyclomatic complexity}~\cite{mccabe:complexity} of a program, defined starting from its control flow graph $\mathcal{G}$ having $n$ vertices, $e$ edges and $p$ connected components as:

\begin{equation}
V(\mathcal{G}) = e - n + p
\end{equation}

Dealing with java methods, such metric can be easily calculated by counting $1$ point for the beginning of the method, $1$ point for each conditional construct and for each case or default block in a switch-case statement, $1$ point for each iterative structure and $1$ point for each Boolean condition. Starting from this software metric, we define the problem as follows:

\begin{problem}
Let $M$ be a set of Java methods, let $c \in \mathbb{N}$, and let $h_c: M \to \{0, 1\}$ be a binary classification rule for the methods. We define, for each $m \in M$:

\begin{equation}
h_c(m) = 
\begin{cases}
0   & \text{if $V(\mathcal{G}_m) < c$} \\
1   & \text{otherwise}
\end{cases}
\end{equation}

\end{problem}

\paragraph{Semantic labelling policy}

For the definition of the semantic labelling strategy, we adopted the same assumption that Allamanis et al.~\cite{allamanis:summarization} and Alon et al.~\cite{alon:pathRep} made in their works, namely that the name a programmer gives to a method can be somehow considered as a summary of the method's operations, meaning that the name of a method shall provide some semantic information on the method itself. Starting from this premise, we define this semantic labelling considering the presence or absence of specific patterns in the method name:

\begin{problem}
Let $N$ be a set of method names, and let $T$ be a set of patterns. We write $r \leq s$ if $r$ and $s$ are strings and $r$ is a substring of $s$. Let $h_T: N \to \{0, 1\}$ be a binary classification rule for the methods. We define, for each $n \in N$:

\begin{equation}
h_T(n) = 
\begin{cases}
1   & \text{if $\exists t \in T \colon t \leq n$}\\
0   & \text{otherwise}
\end{cases}
\end{equation}

\end{problem}
\paragraph{Random labelling policy}

We finally define a baseline labelling strategy for assessing our results by comparing them with the results obtained while solving an arbitrary task whose results should be only noise. We simply consider a random split of the methods:

\begin{problem}
Let $L$ be a randomly shuffled list of methods, and let $m_i \in L$ be the method having index $i$. Given a threshold $n \in \mathbb{N}$ and a function $h_n: L \to \{0, 1\}$, we define, for each $m_i \in L$:

\begin{equation}
h_n(m_i) = 
\begin{cases}
1   & \text{if $i \leq n$}\\
0   & \text{otherwise}
\end{cases}
\end{equation}
\end{problem}

\subsection{Classification}

For assessing the ability of our networks in autonomously building internal representations for diverse program features, we tested the performance of the hidden neurons in classifying methods according to different instances of the classes of problems described in the previous section. To this aim, we considered all the neurons in the code layer and in the two hidden layers in the decoder, as shown in Figure~\ref{fig:autoencoder}, and we tested their classification accuracy by considering the activation produced by a neuron for the method given as input. The reason why we tested only the decoding neurons lies in the nature of an autoencoder: in the encoder layers a progressive dimensionality reduction (and thus a compression of information) is performed, and this (likely) means that in the middle code layer only relevant features are encoded. Since in the decoder layers the dimension is symmetrically increased for reconstructing the input, we decided to test only those neurons, since they are expected to hold more relevant features.

For each of the selected neurons, we considered $10$ equally spaced values among the minimum and the maximum activation value of that neuron for methods in the training set. For each activation threshold, we computed the classification accuracy of the neuron on a given problem instance by considering, in a precomputed balanced sample of the test set, the activation value of the neuron for that method and by predicting the method to be in class $0$ or in class $1$ if the activation value is less or greater than the threshold, respectively. 

\begin{figure}
    \centering
    \includegraphics[width=\linewidth]{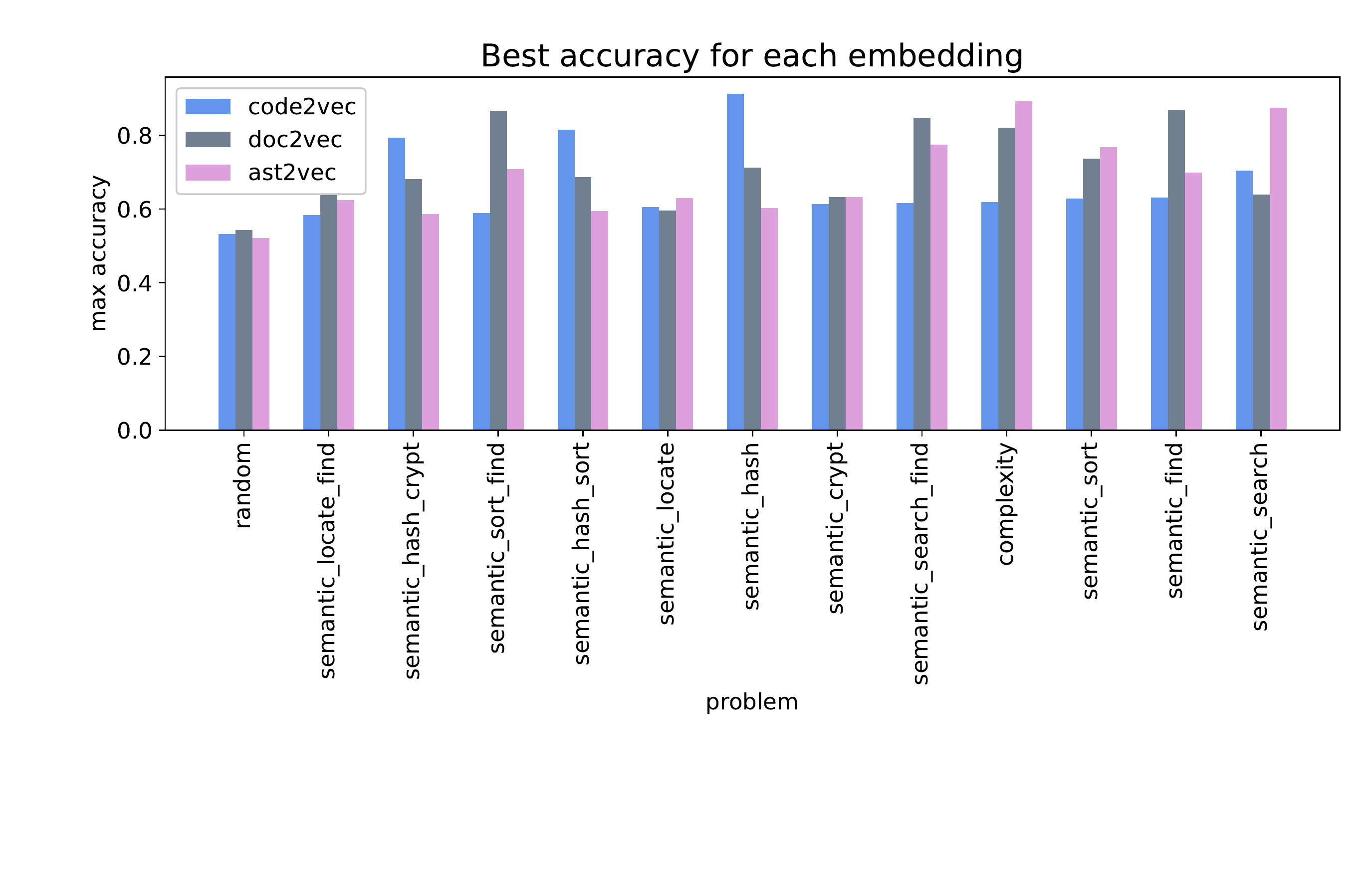}
    \caption{Accuracy obtained by the best neuron for each problem instance.}
    \label{fig:summary}
\end{figure}

This process, formally described in Algorithm~\ref{algo:cls}, gives us a procedure for ranking neurons according to a task: we assign to each neuron its highest accuracy score. Table~\ref{tab:accuracy} and Figure~\ref{fig:summary} show the accuracies obtained by the best neuron for each problem, while the complete results obtained with the classification experiments described above will be discussed with further details in Section~\ref{sec:discussion}.

\begin{figure}
    \begin{algorithmic}[1]
    \State $bestAcc \gets 0$\Comment{accuracy of the best neuron}
        \ForAll{neuron $N$}
            \State $A \gets$ activation values of $N$
            \State $T \gets$ activation thresholds \Comment{$10$ evenly spaced thresholds between $0$ and $\max{a \in A}$}
            \State $best_N \gets 0$\Comment{best accuracy for $N$}
            \ForAll{$t \in T$}
                \State $pred \gets$ empty list\Comment{list of predictions}
                \ForAll{$a \in A$}
                    \If{$a \le t$}
                        \State append $0$ to $pred$
                    \Else 
                        \State append $1$ to $pred$
                    \EndIf    
                \EndFor
                \If{\Call{accuracy}{$pred$} $\ge bestN$}
                    \State $best_N \gets$ \Call{accuracy}{$pred$}\Comment{update best accuracy of $N$}
                \EndIf    
            \EndFor
            \If{$best_N \ge bestAcc$}
                \State $bestAcc \gets best_N$\Comment{update best neuron}
                \State $bestNeuron \gets N$
            \EndIf
        \EndFor
    \end{algorithmic}
    \caption{Algorithm for finding the best neuron in classifying programs on binary problems.}\label{algo:cls}
\end{figure}

\begin{table}
\caption{Best accuracy score for each problem instance}
\label{tab:accuracy}

\begin{center}
\begin{tabular}{llccc}

\hline \\
\textbf{Class}  &\textbf{Instance}          &\textbf{code2vec}  &\textbf{doc2vec}   &\textbf{ast2vec}\\
\hline \\
Random          &none                       &$53\%$             &$54\%$             &$52\%$\\
Structural      &$c=10$                     &$62\%$             &$82\%$             &$89\%$\\
Semantic        &$T=\{\text{sort}\}$        &$63\%$             &$74\%$             &$77\%$\\
Semantic        &$T=\{\text{find}\}$        &$63\%$             &$87\%$             &$70\%$\\
Semantic        &$T=\{\text{search}\}$      &$70\%$             &$64\%$             &$87\%$\\
Semantic        &$T=\{\text{locate}\}$      &$60\%$             &$59\%$             &$63\%$\\
Semantic        &$T=\{\text{sort, find}\}$  &$59\%$             &$87\%$             &$71\%$\\
Semantic        &$T=\{\text{find, search}\}$&$61\%$             &$85\%$             &$77\%$\\
Semantic        &$T=\{\text{find, locate}\}$&$58\%$             &$85\%$             &$62\%$\\
Semantic        &$T=\{\text{hash}\}$        &$91\%$             &$71\%$             &$60\%$\\
Semantic        &$T=\{\text{crypt}\}$       &$61\%$             &$63\%$             &$63\%$\\
Semantic        &$T=\{\text{hash, crypt}\}$ &$79\%$             &$68\%$             &$59\%$\\
Semantic        &$T=\{\text{hash, sort}\}$  &$69\%$             &$69\%$             &$60\%$\\
\hline \\

\end{tabular}
\end{center}
\end{table}

\section{Finding Interesting Neurons}

In the previous section we described our experiments for evaluating the ability of individual neurons in solving specific classification problems or, in other words, in recognising different program properties.

In this section, we discuss two possible approaches for ranking neurons independently from any task. The first approach has been already investigated in the literature (see~\cite{dalvi:sabbia} and~\cite{bau:nmt}) and it uses the correlation between activation values of neurons in distinct but isomorphic models for finding neurons that possibly capture properties that emerge in different models. The second approach aims to find a score measure for neurons based on the concept of \emph{entropy} used in information theory.  
For simplifying the discussion, we performed all the following experiments by considering only the ast2vec embedding, since the classification experiments suggest that it is the most adaptive for this kind of application among the embeddings tested.

\begin{figure*}
    \centering
    \includegraphics[scale=0.39]{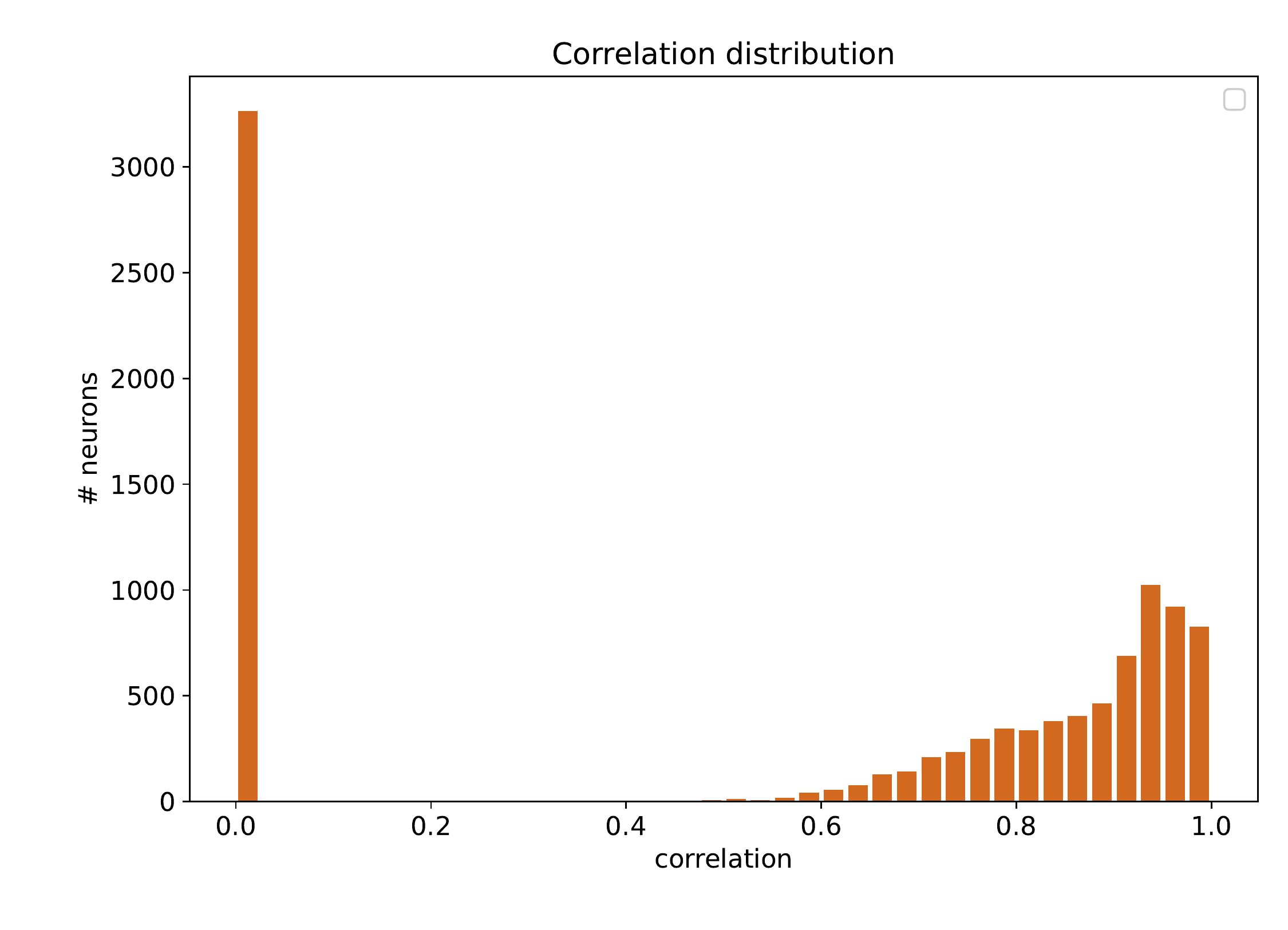}
    \includegraphics[scale=0.39]{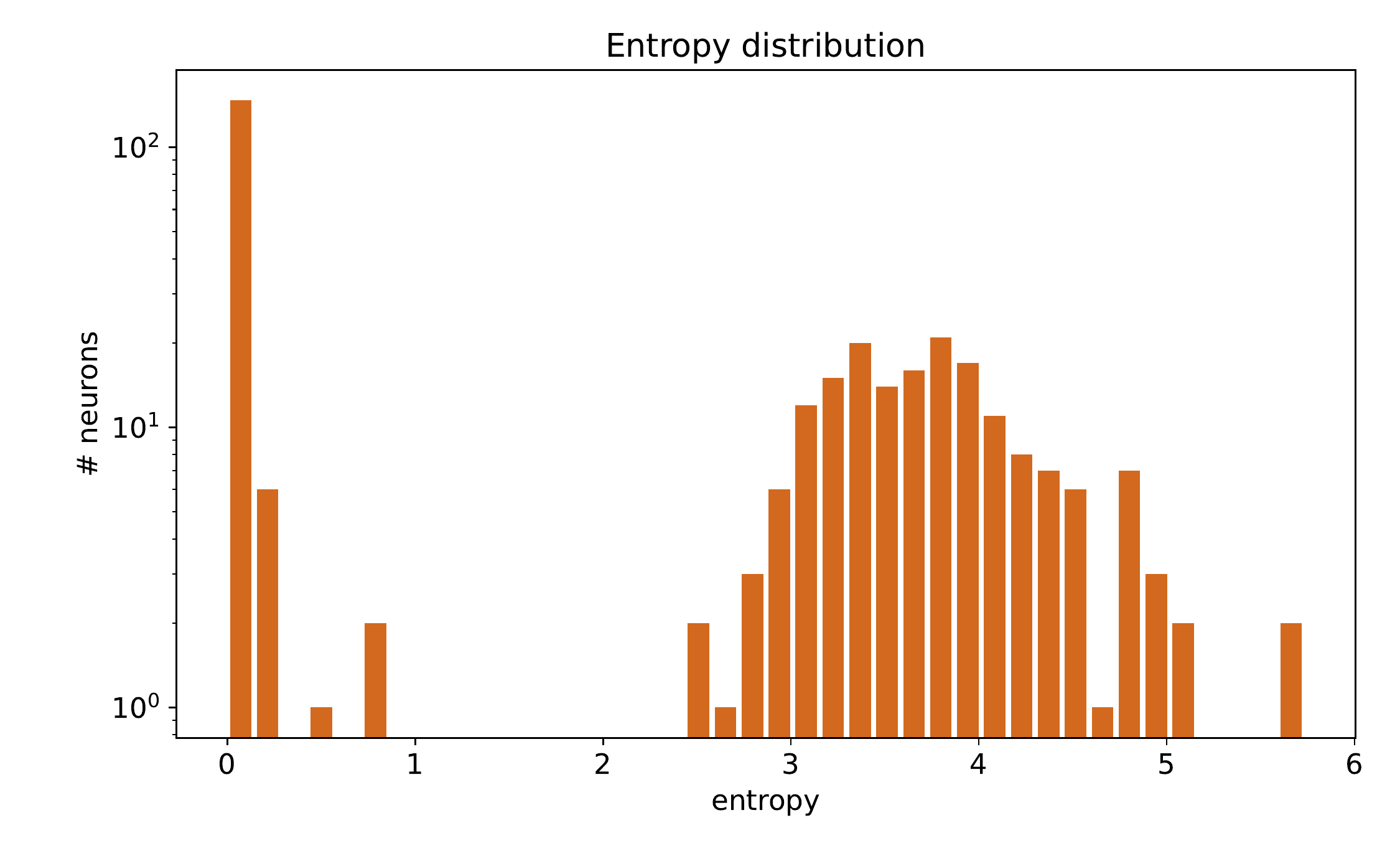}
    \caption{Distribution of the neurons' scores based on the Perason correlation (left) and on the entropy (right). In the correlation approach, the distribution suggests that there aren't groups of neurons that can be significantly distinguished from the others, while interesting entropy values are those between $0.03$ and $1$.}
    \label{fig:regardles}
\end{figure*}

\subsection{Correlation analysis}

The first method we propose for ranking important individual neurons has been already adopted in previous works~\cite{dalvi:sabbia, bau:nmt}. It is based on the search of neurons that in different but similarly trained networks produce similar activation values, and it relies on the strong assumption that single neurons capture the same important properties with respect to the task of the network in different models. Since the task our autoencoers are, by definition, trained on is the reconstruction of the input, the assumption in this context turns to be that important neurons capture important features with respect to the input itself.

We consider $30$ different autoencoders $\mathbb{A}_1, \dots, \mathbb{A}_{30}$ having the same parameters but trained on different vectors, which are selected as a random sample of the full original training set and differ for each model. We refer to the activation values for a test set having cardinality $N$ of the neuron $j$ of the model $i$ as $\mathbb{A}_{i,j} \in \mathbb{R}^N$, and we assign to each neuron a score value $0 \leq s(\mathbb{A}_{i,j}) \leq 1$ defined, for $0 \leq i' \leq 30$ and for $0 \leq j' \leq N$, with $i \neq i'$, as:

\begin{equation}
    s(\mathbb{A}_{i,j}) = \max{\lvert\rho(\mathbb{A}_{i,j}, \mathbb{A}_{i',j'})\rvert}
\end{equation}
where $\rho(\mathbb{A}_{i,j}, \mathbb{A}_{i',j'})$ is the Pearson correlation coefficient. In other words, a neuron reaches an high score if it shares its behaviour (i.e. produce similar patterns of activation values on the same input) with a neuron in another model. 

The distribution of the neurons obtained by considering this score is reported in Figure~\ref{fig:regardles}. As shown in the figure, most of the neurons have a score of $0$ or greater than $0.8$, i.e. almost all the neurons in all the models have a minimal or a very high score. The reason of such high values can be, to some extent, attributable to the fact that some neurons are never (or almost never) active: this causes the neurons having (almost) all the activations equals to $0$ to reach the maximum correlation value, even if these are obviously slightly informative. Moreover,  the remaining neurons all have maximum correlation values between $0.4$ and $0.6$, which are not significant correlation values. Therefore, this scoring method is not suitable in this context for characterizing interesting neurons.   

\subsection{Information theoretic analysis}

The second method we propose for evaluating the importance of each neuron in the network is based on the information theoretical concept of entropy~\cite{shannon:information}. As we will discuss in Section~\ref{sec:discussion}, the experiments performed for assessing the results obtained with this scoring approach proved the potential of this ranking, since it can clearly discriminate between important and unnecessary neurons. Moreover, we are able to characterize a class of neurons that are able to detect methods sharing similar properties, which makes this approach particularly suitable for many applications and further research in the program comprehension field.       

The baseline idea beyond this approach is that each neuron can be seen as a signaling system whose symbols are its activation values. Formally, in information theory the entropy is defined as the average information obtained from a signaling system $S$ which can output $q$ different symbols $s_1, \dots, s_q$ with probability $p_i = P(s_i)$:

\begin{equation}
    H(S) = \sum_{i = 0}^{i \leq q}{p_i \log{\frac{1}{p_i}}}
\end{equation}
Dealing with activation values, whose domain is continuous over $\mathbb{R}^+_0$, we constructed a discretization of that space by considering a set $R = \{r_1, \dots, r_{1000}\}$ of $1000$ evenly spaced intervals between $0$ and the maximum activation value reached by a neuron for the vectors in the training set, and we considered those intervals as the possible symbols of the neurons' alphabet. More precisely, for each neuron $N$ we computed the activation values yielded on a random sample of $10000$ vectors, and we determined the number of occurrences of each symbol by counting the activation values in each interval $r_i$. We then considered the set $R_N \subseteq R$ of the occurring symbols in the neuron $N$ and for each $r_i \in R_N$ we derived its occurring probability $p_i$ using a softmax function over the set of countings. Then, we assigned to each neuron a score defined as the entropy computed over the set $P_N = \bigcup p_i$, as described in Algorithm~\ref{algo:entropy}. 

\begin{figure}
    \begin{algorithmic}[1]
    \State $R \gets$ list of intervals 
    \ForAll{neuron $N$}
        \State $M \gets$ random sample of $10000$ methods
        \State $V \gets$ activations of $N$ for each $m \in M$ 
        \State \Call{score_neuron}{$N, R, V$}
    \EndFor
    \Procedure{score_neuron}{$N, R, V$}
        \State $C \gets$ empty list
        \ForAll{$r \in R$}
            \State $c \gets$ number of $v \in V$ such that $v \in r$
            \State append $c$ to $C$
        \EndFor
        \State remove all the $0$s from $C$
        \State $P \gets $ \Call{softmax}{$C$}
        \State \textbf{return} $ -\sum_{p_i \in P}{p_i \log{p_i}}$
    \EndProcedure
    \end{algorithmic}
    \caption{Algorithm for computing the entropy of each neuron.}\label{algo:entropy}
\end{figure}

As is evident in Figure~\ref{fig:regardles}, which reports the entropy values distribution among the neurons, three classes of neurons having different entropy values can be distinguished: 

\begin{enumerate}
    \item A big number of neurons (notice that the figure is in a logarithmic scale) having an entropy equals or very close to $0$. These neurons (which are exactly the same that reach a maximal correlation-based score) are of no interest in this context, since they are neurons that (almost) never activate. They could only be used for optimising the network architecture but it is out of the focus of this work.
    
    \item Another big class of neurons having normally distributed high entropy values. Those neurons reach an high score since their activation values are distributed over a wide range. In addition, the probabilities of the occurring activation values to be in distinct intervals are relatively similar: this leads to an high score in terms of information theory, but makes such neurons barely useful for characterising program properties or interesting behaviours.
    
    \item A smaller set of neurons having values higher than $0$ but less than $1$. Such neurons have been found to be the most interesting for our purposes, as it will be clear by the discussion in Section~\ref{sec:discussion}. Those neurons are peculiar, since they produce an activation higher than $0$ only for some sets vectors, while in all the other cases their activation is equal to $0$  
\end{enumerate}
The behaviour of the third class of neurons is particularly interesting, since it allows us to clearly distinguish classes of methods that prominently stimulate a given neurons, suggesting that such methods share a particular property that can be characterized.

\section{Discussion}\label{sec:discussion}

In this section, we discuss the results obtained with our experiments. We first analyse in details the results obtained in the classification experiments, pointing out the strength of our promising analysis approach, while in the second part of this section we describe and comment the experiments performed for studying the behaviour of neurons that the entropy-based scoring revealed to be the most interesting.

\subsection{Classification results}

As introduced in Section~\ref{sec:classification}, we trained our autoencoders in reconstructing three kinds of program vectors: the task-dependent code2vec embedding, the embedding obtained by the simple application of the doc2vec model on the source code and the ast2vec embedding which is built by considering both syntactical and lexical information. We then fed as input for inference different sets of vectors consisting in balanced classes of methods labelled according to different classification rules, each representing a particular program property.  

\begin{figure}
    \centering
    \includegraphics[width=\linewidth]{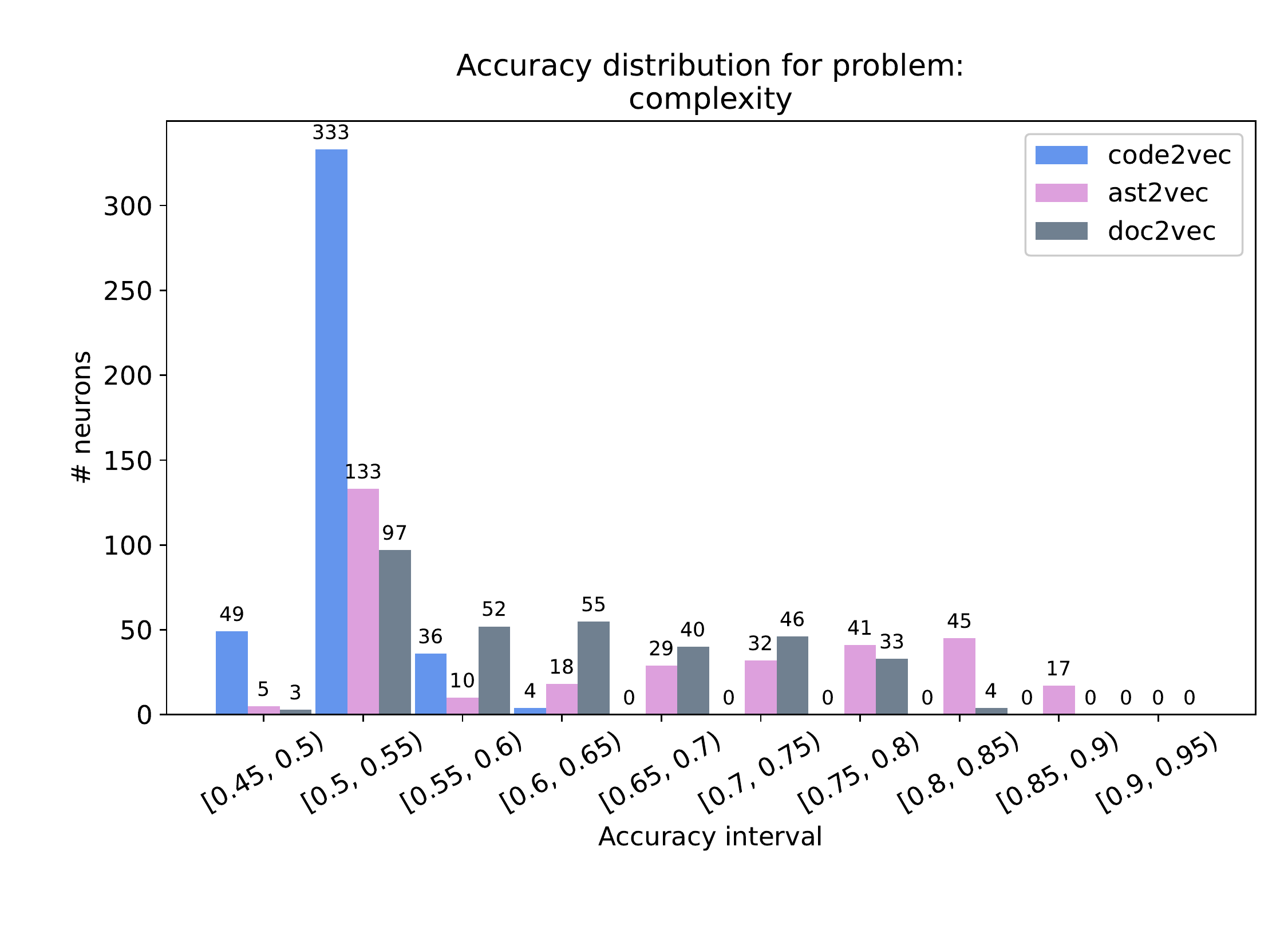}
    \caption{Accuracy distribution obtained in the structural problem. The embedding reaching the highest accuracy values is ast2vec, since it is built starting from the AST.}
    \label{fig:struct}
\end{figure}

\begin{figure}
    \centering
    \includegraphics[width=\linewidth]{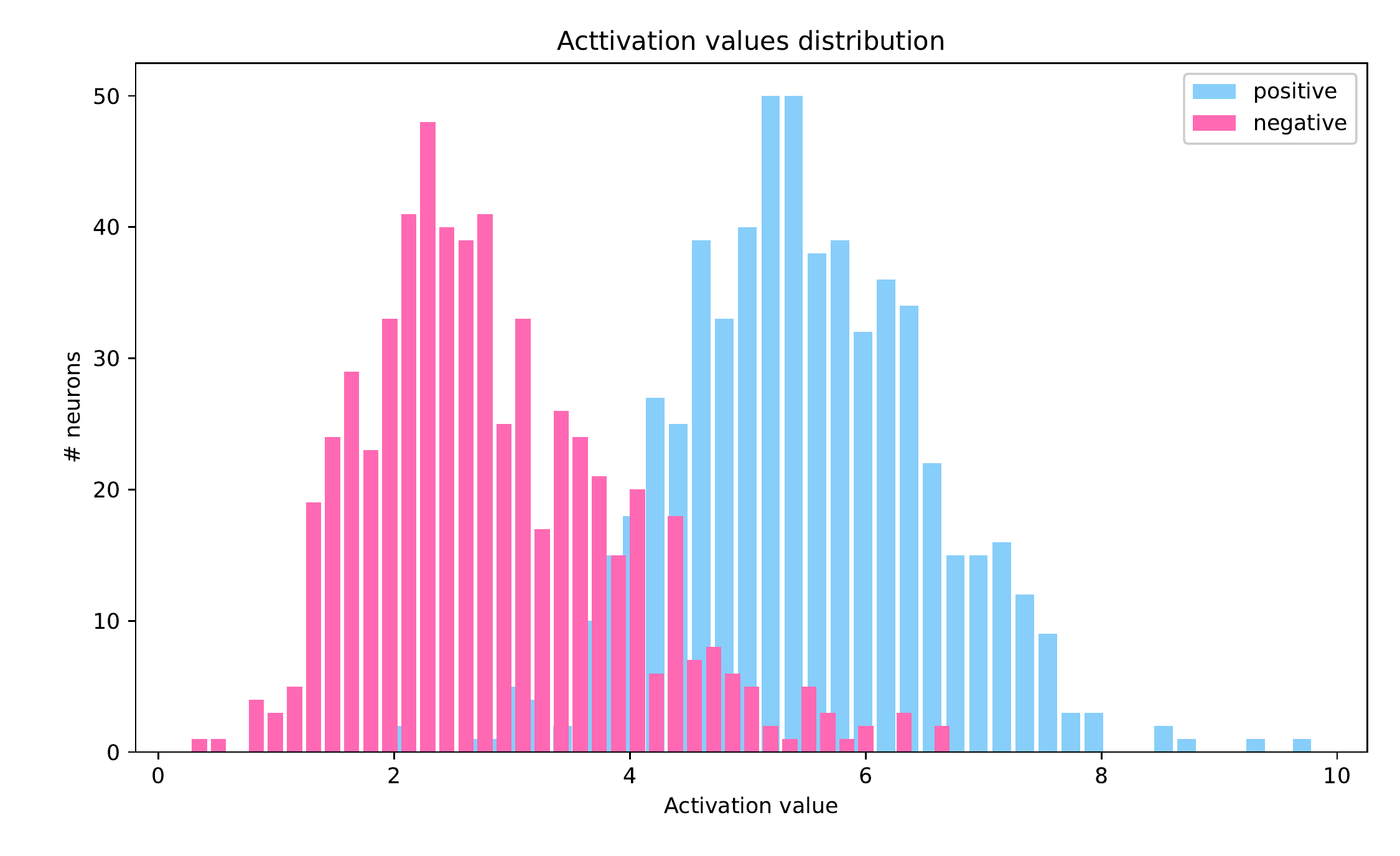}
    \caption{Distribution of activation values of the best neuron in the structural problem.}
    \label{fig:actval}
\end{figure}

The first classification experiment, defined starting from the cyclomatic complexity in order to consider structural features, revealed an accuracy score (see Table~\ref{tab:accuracy}) of $0.89$ reached by a neuron in the ast2vec autoencoder. Furthermore, as reported in Figure~\ref{fig:struct}, the distribution of the accuracies shows that the neurons ast2vec autoencoder reach, in general, an higher accuracy than those of other autoencoders. This should not surprise: since ast2vec embedding is built by considering syntactical features of programs, it is expected to hold in itself the structural information needed for solving this particular task. As an additional evidence of this reliability,  Figure~\ref{fig:actval} shows that activation values for this problem have distinct distributions for the vectors belonging to different classes.

\begin{figure}
    \centering
    \includegraphics[width=\linewidth]{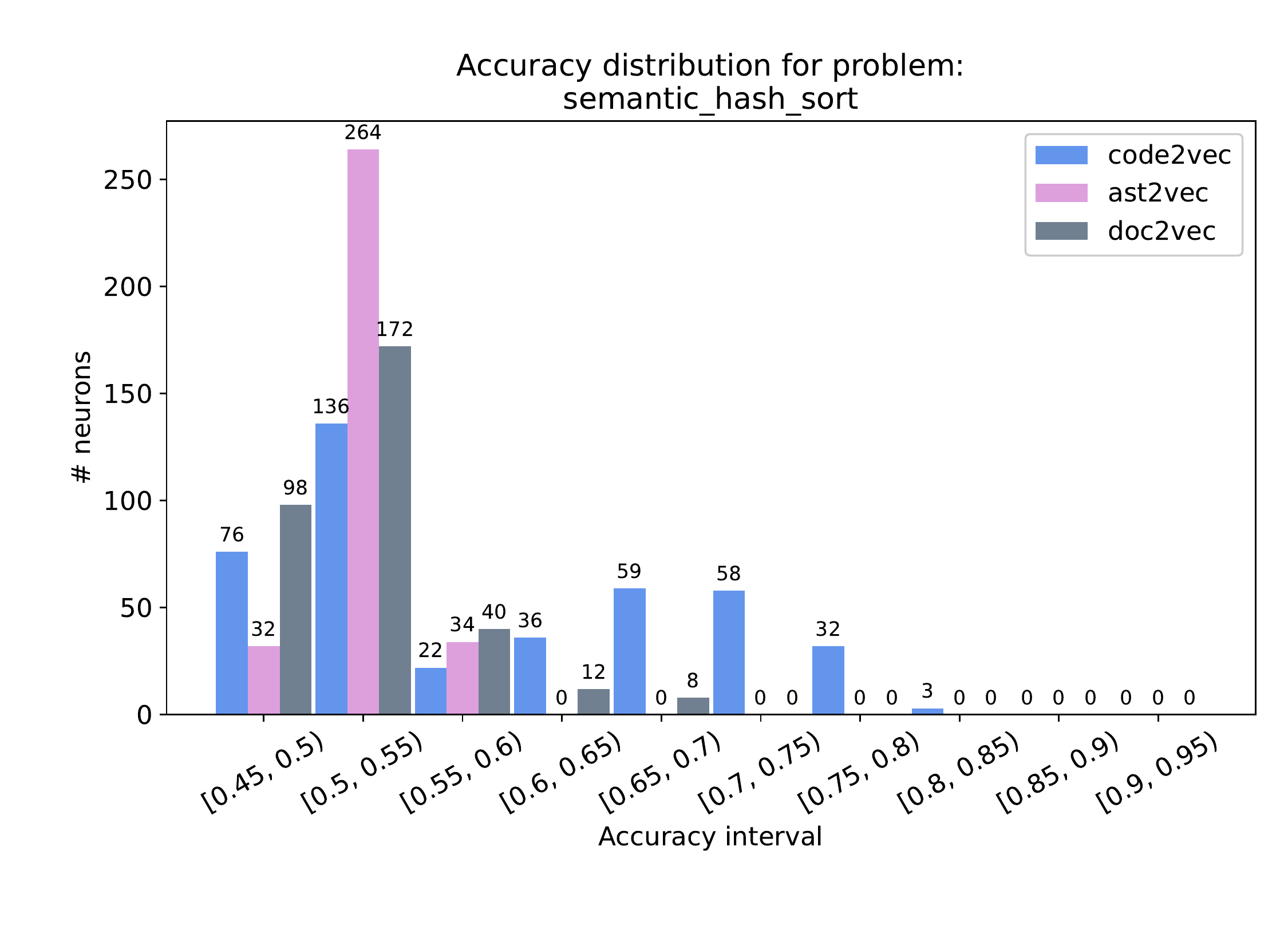}
    \caption{Accuracy distribution obtained in the semantic_hash_sort problem.}
    \label{fig:hashsort}
\end{figure}

\begin{figure*}
    \centering
    \includegraphics[scale=0.39]{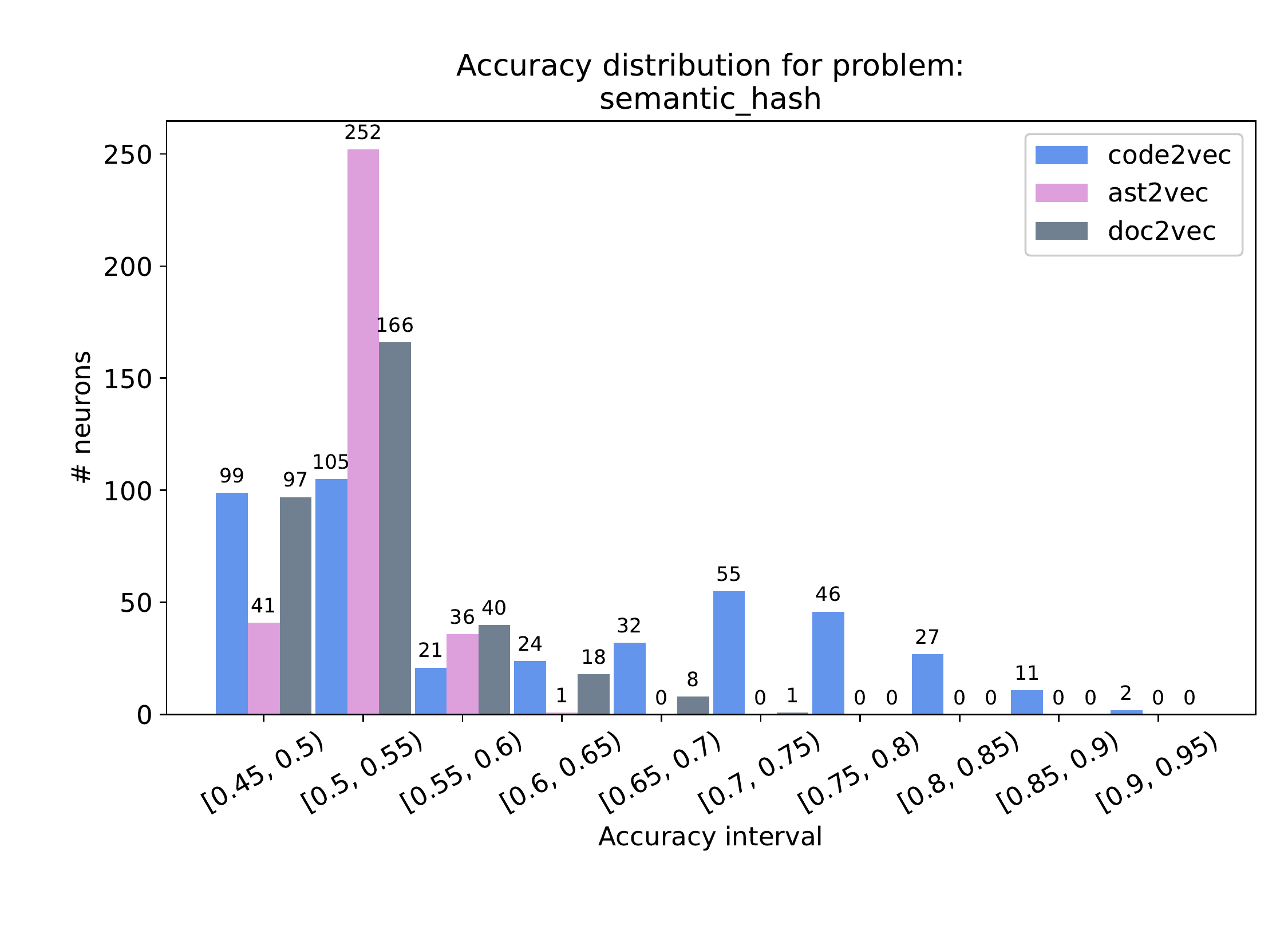}
    \includegraphics[scale=0.39]{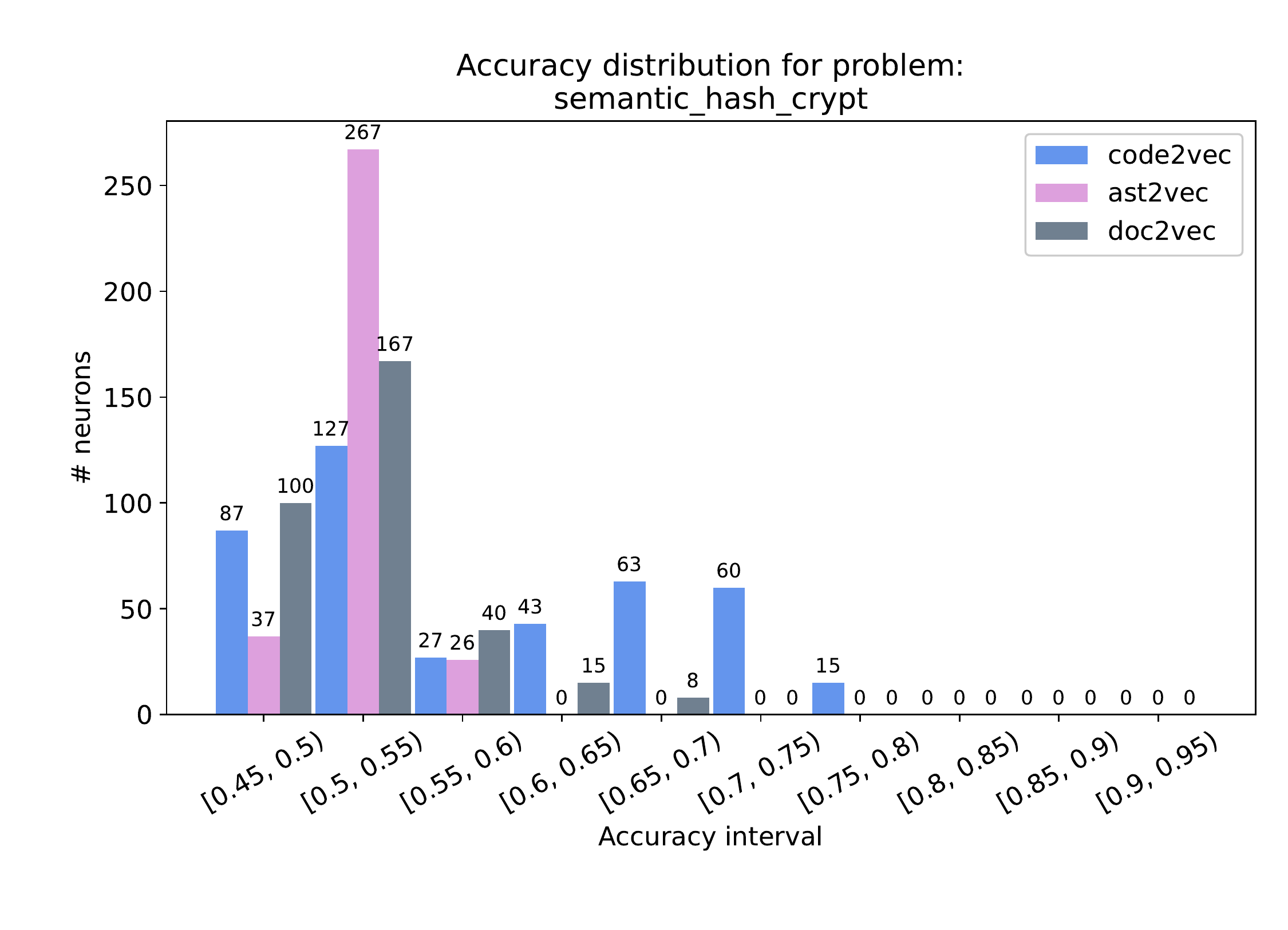}
    \caption{Accuracy distribution obtained in the semantic_hash (left) and semantic_crypt (right) problems.}
    \label{fig:crypto}
\end{figure*}

For what concerns the semantic problems, we decided to test our models on different, relevant instances. We selected terms that represent different operations or contexts. In the problem instances belonging to the cryptographic field (Figure~\ref{fig:crypto}), neurons in the code2vec autoencoders turn out to be the best in classifying methods containing in their name the word \emph{hash}: they reach for this problem an accuracy of $0.91$. We also considered the term \emph{crypt} (aiming to include all the methods concerning encryption and decryption operations) and in all the autoencoders the best neurons reach an accuracy of about $0.6$.   

We also tested our neurons in classifying methods according to the operations of sorting and searching. The terms we considered for these experiments are \emph{sort}, \emph{find}, \emph{search} and \emph{locate}. In methods related to these concepts (Table~\ref{tab:accuracy} and Figure~\ref{fig:summary}, neurons of different autoencoders result to be the most accurate depending on the problem instance. 

\begin{figure*}
    \centering
    \includegraphics[scale=0.39]{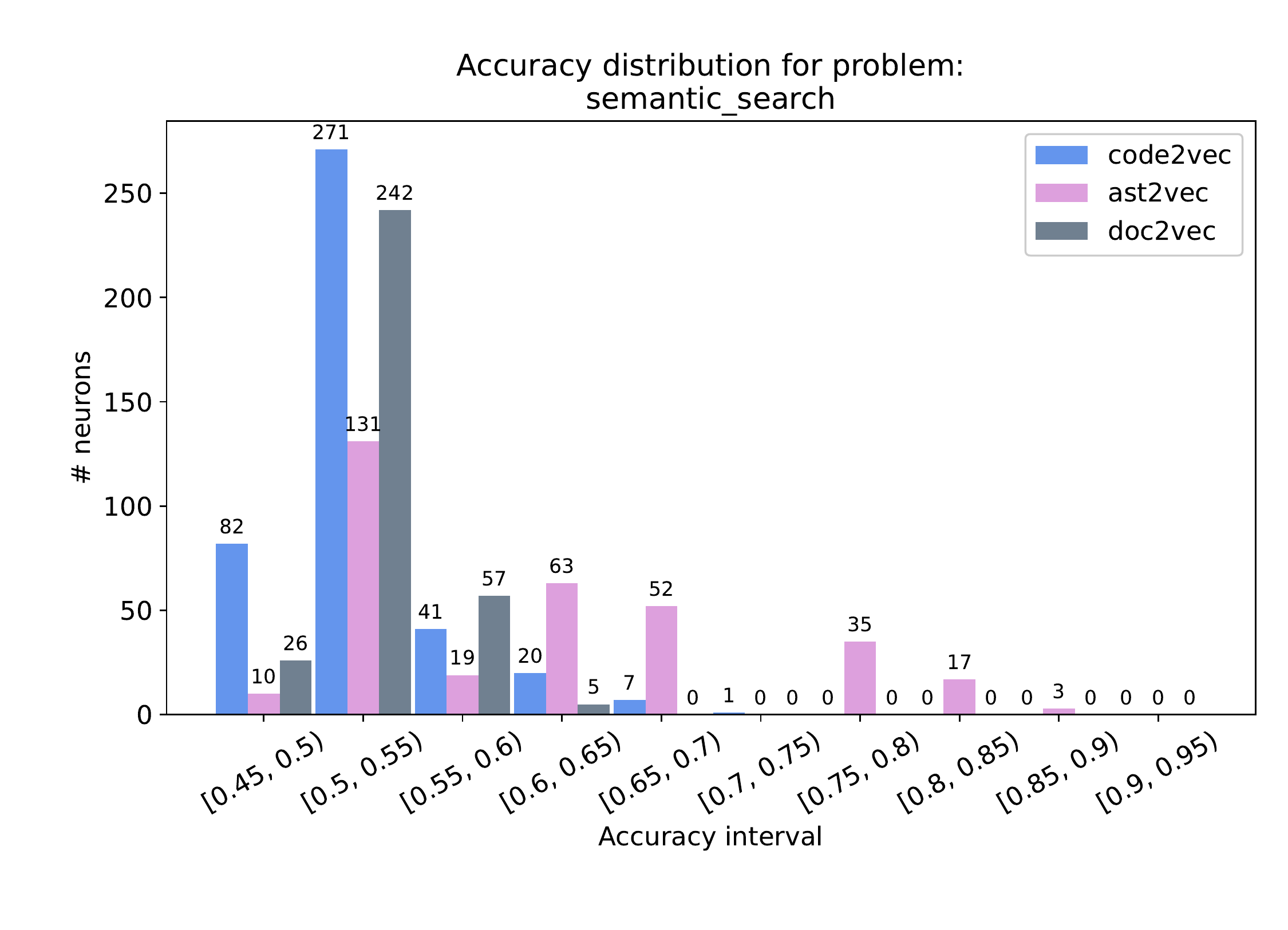}
    \includegraphics[scale=0.39]{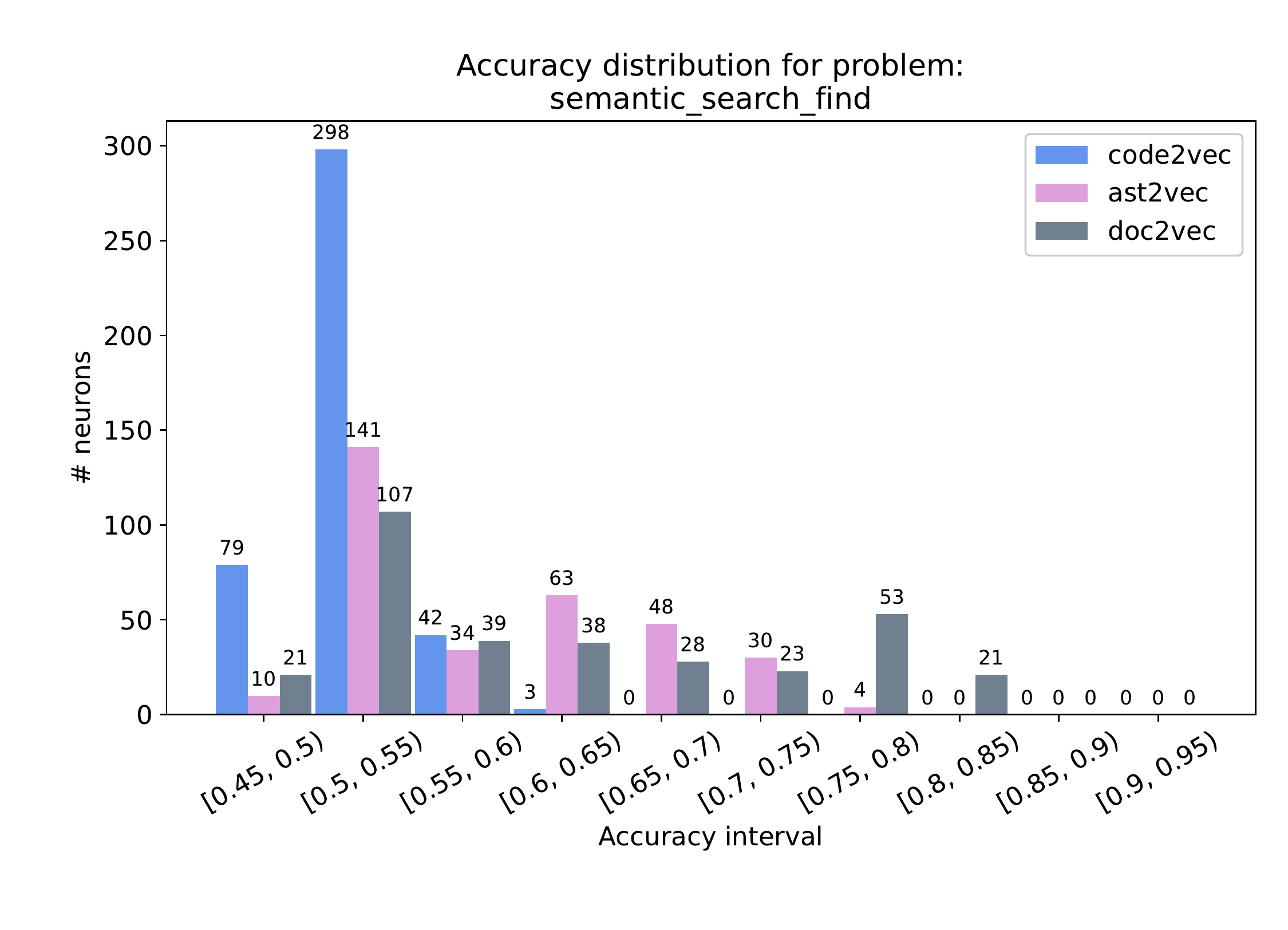}
    \caption{Accuracy distribution obtained in the semantic_search (left) and semantic_search_find (right) problems.}
    \label{fig:find}
\end{figure*}

For a more accurate assessment of the concepts captured by the neurons, we even considered combinations of terms, both belonging to the same semantic domain (for instance \emph{hash} and \emph{crypt} or \emph{find} and \emph{locate}) and to different ones (e.g. \emph{hash} and \emph{sort}, Figure~\ref{fig:hashsort}).

In general, we can say that, except for the problems where the term \emph{hash} is considered, where code2vec neurons result to be always the best, doc2vec and ast2vec neurons are alternating in being the most accurate, see for example Figure~\ref{fig:find}. 

\begin{figure}
    \centering
    \includegraphics[width=\linewidth]{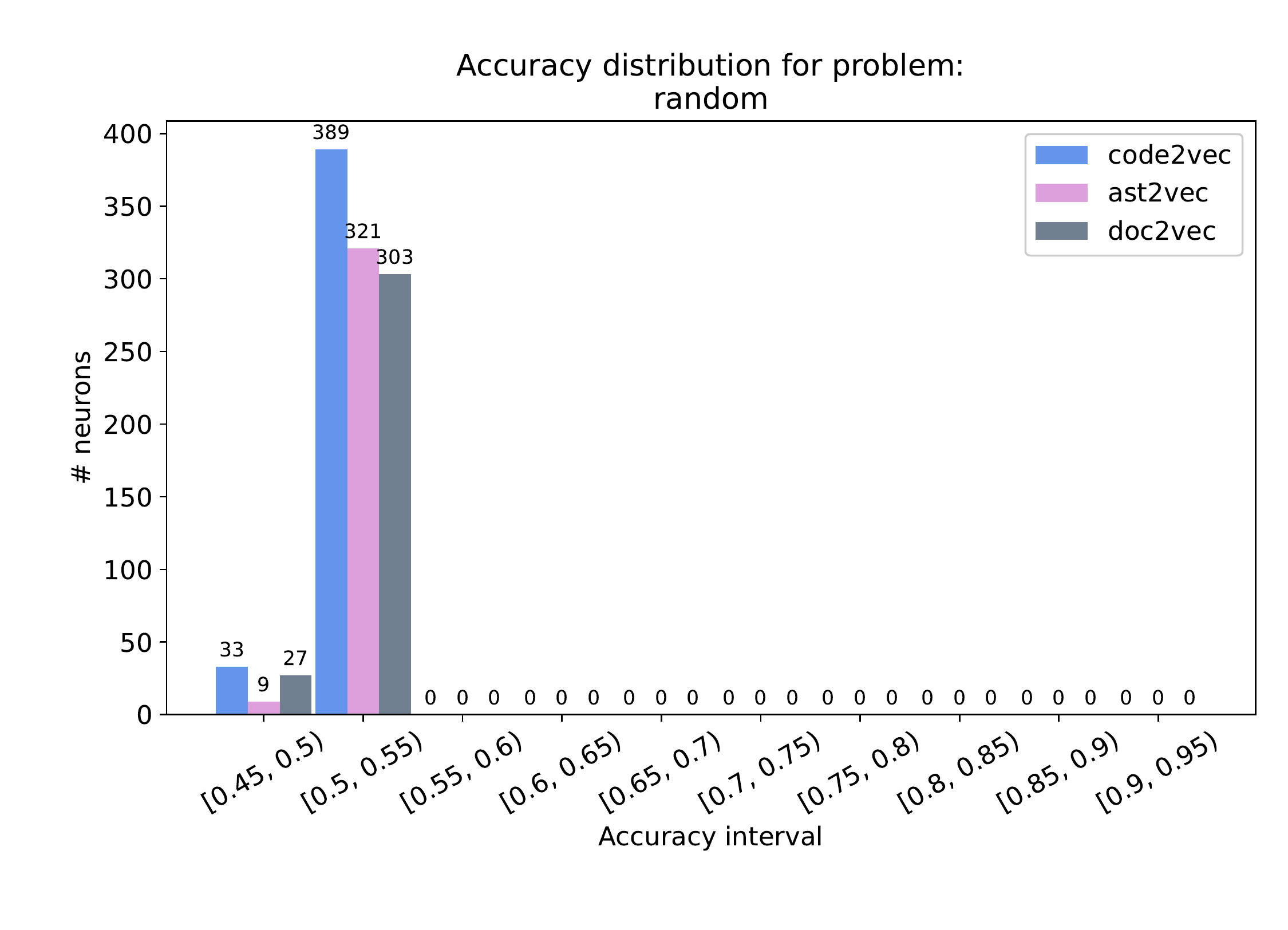}
    \caption{Accuracy distribution on the baseline problem defined by random labelling. Almost all the neurons reach an accuracy level between $0.50$ and $0.55$, showing that accuracies reached in the other problems are all reasonably higher than one reached in a \enquote{lucky} random problem.}
    \label{fig:random}
\end{figure}

Finally, we find useful to visualize the overall behaviour of our systems on the random problem. Figure~\ref{fig:random} shows how all the neurons of all the autoencoders reach an accuracy score near to $0.5$: this states the significance of our results. The accuracy reached in all the other problems is always significantly higher than the maximal accuracy obtained in this random problem: this allow us to clearly differentiate the results obtained on the duly defined problems from those obtained on a possible \enquote{lucky} randomly defined one.  

\subsection{Entropy-based ranking}

Our entropy-based scoring has been applied to each neuron of the autoencoder layers highlighted in Figure~\ref{fig:autoencoder}. Those three layers are expected to have different dynamics, where the sparseness of activation levels changes from the smallest layer, mostly acting as encoder, to the largest one, contirbuting to the output decoder. This is relevant, since we compute the score by first looking at the overall range of activation values occurring in all these layers, to define common reference range, and then by sampling the activation levels of each neuron in its own dynamics. Therefore we were ready to find for encoding neurons entropy scores different from those of decoding neurons. Secondly, we looked for neurons with low entropy, since we consider that such neurons are able to produce activation levels either fixed, for many input methods, or with variable values for a small portion of the input methods.

And this was in fact what we found out during the experiments on entropy-based scoring. First of all, each neuron is producing activation values that we group by intervals, as previously explained, and this amounts to measure entropy on different sets of symbols for each neuron. After normalizing entropy values according to maximal attainable level given the cardinality of the symbol set of each neuron, we found out that the smaller layers showed high entropy levels more frequently. From this, we moved to selecting neurons we considered interesting for their low entropy.

\begin{figure}
    \centering
    \includegraphics[width=\linewidth]{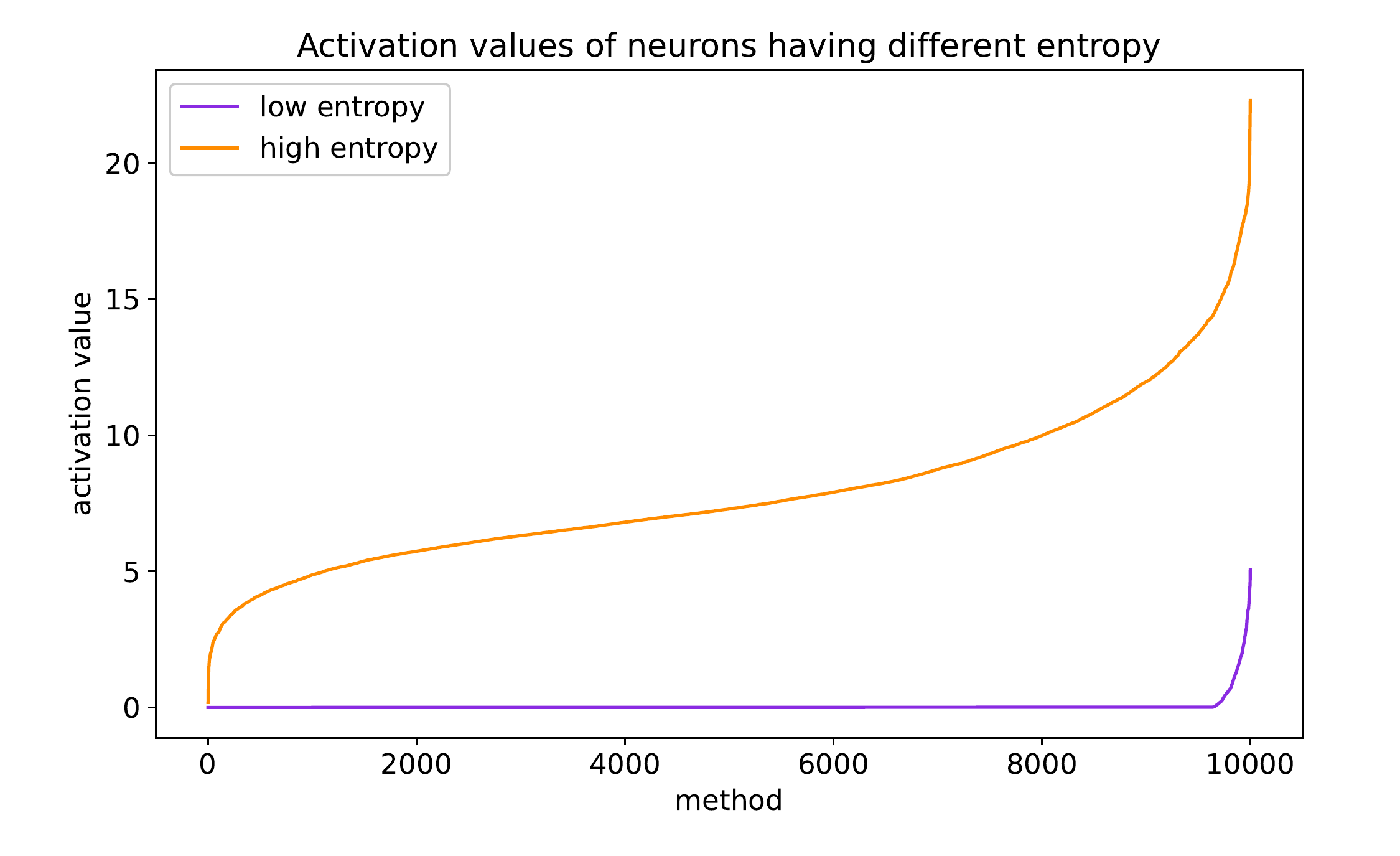}
    \caption{Activation values for a sample of $10000$ methods of neurons having entropy equals to $0.3$ (purple line) and $5.8$ (orange line).}
    \label{fig:activations}
\end{figure}

Eventually, we came to chose one neuron from the middle layer, of the three we focused on, and eight from the largest layer. We checked the distribution of their activation values, produced for each input methods, and we could compare their output range to that of high entropy neurons, with results such as that of Figure~\ref{fig:activations}, showing how our scoring method distinguishes the different distributions of activation values, as expected from a measure inspired by information-theoretic entropy.

The final step was to look for evidence of how the small portion of methods inducing high activation levels, in the neurons we selected as interesting, could be considered as sharing some property. Ideally, the training of the network led those neurons to be able to perform some specific classification of input methods.

\begin{figure}
    \centering
    \includegraphics[width=\linewidth]{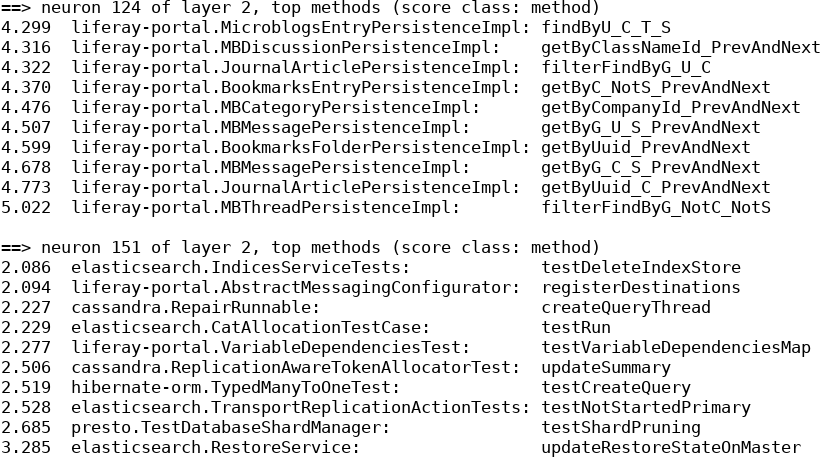}
    \caption{Examples of top methods according to low entropy neurons.}
    \label{fig:cosa}
\end{figure}

The results of this last check on the performance of our scoring method were actually good: we traced back to the input methods which produced high activation levels from the neurons we had selected (the rightmost part of the lower line in Figure~\ref{fig:activations}), and we tried to understand which properties led them to be clustered by each neuron.
We could not come to an explicit definition of such properties, but we consider as evidence that those clusters are not just random groups by looking for instance to the results showed by Figure~\ref{fig:cosa}. There we can see that ``top methods'' of the first classifying cluster, induced by activation levels of neuron 148 in second layer, share some property related to their goal, as hinted by methods' names, and also are all in the same repository.

On the other hand, the second group of methods in Figure~\ref{fig:activations} could be associated on some property difficult to explicate, by just considering methods' names. We also looked at the actual source code of those methods, but we could not recognize any apparent shared property grouping them. This is search for classifying properties of each cluster will be considered in further work and on more sophisticated than the simple testbed we chose here.

\section{Conclusion and further directions}

In this work, we proposed and investigated a novel approach for source code analysis, based on the study of the behaviour of the internal neurons of artificial neural networks trained on vectors representing snippets of code (Java methods in our experiments, but could be generic parts of source code, no matter the granularity). The results point out the considerable potential of our method, and suggest many directions for further research.

The classification experiments suggest that the embedding function used for preparing the input could significantly affect the performance. Our results show that ast2vec and doc2vec vectors are more suitable than code2vec for these general applications, but different code vectors deserve to be tested. As an example, cuBERT contextual vectors\cite{kanade:cubert} could be particularly good for these general comprehension analysis, but also customized new embeddings could be designed in order to make the input to include in itself relevant features that could be exploited by the network for building internal representations of high-level concepts. 

Our information theory-based method for ranking neurons independently from any task is also promising. Our analysis shows that different neurons in a network are tending to be specialized in different functions. In our context, single neurons in the middle layer which share, in general, higher entropy scores, probably do not encode program features by themselves alone, while the more external ones seem to be more sensitive to specific properties, as suggests the fact that some of them only activate on particular sets of methods. A deeper analysis of such methods could allow us to characterize them, and this characterization could be used for define new, relevant, program properties. 

Many applications are possible for exploiting the insights described in this work. Neural models (also more sophisticated and optimized than the simple autoencoders used here) can be embedded in different environments for easing the cognitive effort of the programmers or for helping users in their programming tasks. As an example, we suggest applications in software repositories or text editors. The capability of the neurons in recognising specific semantic properties can be used for searching code using queries expressed in natural language or it can be also applied for producing an automatic tagging of programs or projects. Such models can be also included in editors or IDEs for highlighting programming good/bad practices, possible bugs, or for assisting in the learning of new programming languages.  

\bibliographystyle{IEEEtran}
\bibliography{main.bib}

\end{document}